\newcommand{\solar}{L$_{\odot}$\ }
\newcommand{\solm}{M$_{\odot}$}

\documentclass{iau}
\usepackage{graphicx}

\title[NIR K-band identification of DSO/G2]{The infrared K-band identification of the DSO/G2 source from VLT and Keck data}

\author[Andreas Eckart et al.]  
{
A. Eckart$^{1,2}$,
M. Horrobin$^{1}$,
S. Britzen$^{2}$, 
M. Zamaninasab$^{2}$, 
K. Mu\v{z}i\'{c}$^{3}$, 
N. Sabha$^{1,2}$, 
B. Shahzamanian$^{1,2}$,
S. Yazici$^{1}$, 
L. Moser$^{1}$, 
M. Garcia-Marin$^{1}$, 
M. Valencia-S.$^{1}$, 
A. Borkar$^{2,1}$, 
M. Bursa$^{4}$, 
G. Karssen$^{1}$, 
V. Karas$^{4}$, 
M. Zaja\v{c}ek$^{4,5}$,
L. Bronfman$^{6}$,
R. Finger$^{6}$,
B. Jalali$^{1}$, 
M. Vitale$^{2,1}$,
C. Rauch$^{2}$, 
D. Kunneriath$^{4}$, 
J. Moultaka$^{7,8}$,
C. Straubmeier$^{1}$,
Y.E., Rashed$^{1}$,
K. Markakis$^{2,1}$, 
\and
A. Zensus$^{2,1}$
 }

\affiliation{
1) I. Physikalisches Institut, Universit\"at zu K\"oln,
           Z\"ulpicher Str. 77,
           50937 K\"oln, Germany
\\
2)            Max-Planck-Institut f\"ur Radioastronomie, 
           Auf dem H\"ugel 69, 
	   53121 Bonn, Germany
\\
3)            ESO, Alonso de Cordova 3107, Vitacura, Casilla 19, Santiago, 19001, Chile 
\\
4)           Astronomical Institute, Academy of Sciences, CZ-14131 Prague, Czech Republic
\\
5)        Faculty of Mathematics and Physics, CZ-18000 Prague, Czech Republic
\\
6)            Departemento de Astronomia, Universidad de Chile, Castilla 36-D, Santiago, Chile
\\
7)            Universit\'e de Toulouse; UPS-OMP; IRAP; Toulouse, France
\\
8)            CNRS; IRAP; 14, avenue Edouard Belin, F-31400 Toulouse, France
}

\pubyear{2013}
\volume{303}  
\pagerange{001--004}
\jname{The GC: Feeding and Feedback in a Normal Galactic Nucleus}
\editors{Editors}
\begin{document}

\maketitle

\begin{abstract}
A fast moving infrared excess source (G2) which is widely interpreted
as a core-less gas and dust cloud approaches Sagittarius~A* (SgrA*) on a presumably
elliptical orbit.
VLT K$_s$-band and Keck K'-band data result in clear continuum identifications and proper
motions of this $\sim$19$^m$ Dusty S-cluster Object (DSO).
In 2002-2007 it is confused with the star S63, but free of confusion again since 2007.
Its near-infrared (NIR) colors and a comparison to other sources in the field
speak in favor of the DSO being an IR excess star with photospheric continuum
emission at 2 microns than a core-less gas and dust cloud.
We also find very compact L'-band emission ($<$0.1'') contrasted by the
reported extended (0.03'' up to $\sim$0.2'' for the tail)  Br$\gamma$ emission.
The presence of a star will change the expected accretion phenomena,
since a stellar Roche lobe may retain a fraction of the material during and 
after the peri-bothron passage.
\end{abstract}

\firstsection 

\vspace*{.5cm}
SgrA* at the center of our galaxy is associated with a
$4 \times 10^{6}$\solar super-massive central black hole.
It is a highly variable radio, near-infrared (NIR) and X-ray source.
In early 2014 the dusty G2/DSO object (Gillessen et al. 2012ab, Eckart et al. 2013b)
will pass by SgrA* at a distance between 120 and 200 AU (1500 and 2400 Schwarzschild radii;
Phifer et al. 2013, Gillessen 2013b)
and is expected to loose matter or even be completely disrupted 
during the periapse section of its orbit - possibly resulting in quite 
luminous accretion events.
We expect that the NIR/X-ray flux density of SgrA* will increase substantially.
The enhanced activity will be strong in the mm- and sub-mm part as well.
To probe the accretion process and investigate geometrical aspects 
(outflow and disk orientation) of the enhanced activity,
(sub-)millimeter/radio observations between June 2013 and (beyond) April 2014
in parallel with the NIR polarization observations will be essential.
So far our NIR VLT, sub-mm APEX and mm ATCA monitoring runs in June, August and September
did not show any exceptional flux density variations.
The activity, however, expected in 2014
will also give an outstanding opportunity to improve the derivation of the spin and
inclination of the SMBH from NIR/mm observations.

\begin{figure}[b]
\begin{center}
\vspace*{-0.25 cm}
 \includegraphics[width=3.4in]{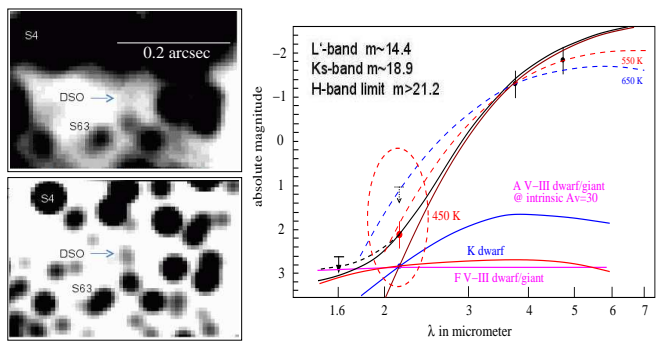} 
 \caption{{\bf Left:} Identification of the DSO/G2 source in the raw and deconvolved Keck images
from August 2010. Top: raw adaptive optics images; Bottom: deconvolved with a 
PSF extracted from the image and re-convolved with a Gaussian to an angular resolution
close to that achieved during these observations.
{\bf Right:} Decomposition of the DSO spectrum including our K$_s$-band detection and H-band limit. 
A mixture of dust and stellar contribution is possible.
The dashed ellipse highlights the 2$\mu$m limit (Gillessen et al. 2013ab) and our detection.
}
   \label{fig1}
\end{center}
\end{figure}

Gillessen et al. (2012ab) interpret the G2/DSO source as a preferentially core-less gas and dust cloud
approaching SgrA* on an elliptical orbit.
Eckart et al. (2013a) present the first K$_s$-band identifications
and proper motions of the DSO.
As further support, Eckart et al. (2013b) present the results of the analysis of
4 epoch of public Keck K'-band adaptive optics imaging data (2008 to 2011; see Tab.1 in
Eckart et al. 2013b).
Based on the comparison to VLT NACO L'- and K$_s$-band data (Eckart et al. 2013ab, Gillessen et al. 2013) 
we can clearly identify the DSO in its K'-band continuum emission as measured by the NIRC2 camera 
at the Keck telescope.
The G2/DSO counterpart can even be seen in the direct (not deconvolved) 
Keck adaptive optics data (Fig.\,\ref{fig1}, left).
For all four public Keck epochs (2008, 2009, 2010, 2011) very similar structures compared 
to those derived from the VLT data presented by Eckart et al. (2013a).

The  NIR colors of the DSO imply that it could rather be an IR excess star.
Very compact L'-band emission is found (pointing at the presence of a star), 
contrasted by the broad
Br$\gamma$ emission (pointing at the presence of a very extended 
optically thin tail or envelope) reported by 
Gillessen et al. (2012ab) and modeled by Burkert et al. (2012) and Schartmann et al. (2012).
The presence of a star will change the expected accretion phenomena (observable through 
expected excess mm- NIR and X-ray flux) since a stellar Roche lobe may retain much of the 
material (Eckart et al. 2013ab) during and after the peri-bothron passage.

\vspace*{-0.45 cm}
\begin{figure}[b]
\begin{center}
 \includegraphics[width=3.4in]{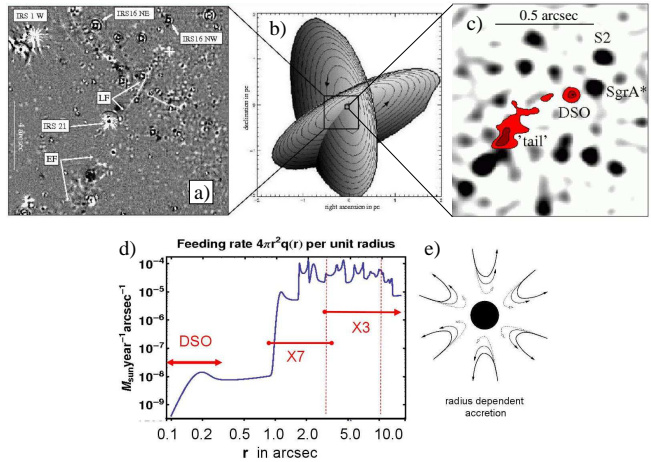} 
 \caption{
{\bf a-c)} The orientation of the G2-tail emission 
(Gillessen et al. 2013)
with respect to
other features at similar position angles like
the interaction zone of the disk system associated with the 
mini-spiral; {\bf b)} Vollmer \& Duschl (2000)
and a dust filament visible in high-pass filtered L-band images (in {\bf c)}).
The linear feature LF crosses the northern arm
and the extended feature EF is associated with the eastern arm
(Eckart et al. 2006).
{\bf d)} Mass input into the feeding region around the SgrA* black hole. 
Using square averaged wind velocities feeding is averaged over stellar orbits (Shcherbakov \& Baganoff, 2010). 
The approximate distances of the DSO and the two cometary shaped sources X3 and X7 (Muzic, Eckart, Sch\"odel et al. 2007, 2010)
are shown. {\bf e)} Sketch of the radius dependent accretion onto the central black hole. Only a very small fraction of the
matter that can be accreted reached the center. A dominant fraction of it is blown away in an outward-bound 
wind component.
}
\label{fig2}
\end{center}
\end{figure}

\section{\bf The Broadband spectrum of the DSO}
In Fig.~\ref{fig1}(right) we show a possible spectral decomposition 
(Eckart et al. 2013a)
of the DSO using the M-band measurement by Gillessen et al. (2012a).
The interesting arising question is about the nature of the G2/DSO K-band emission.
Could the K$_s$-band flux density be due to emission from warm dust?
Most of the mini-spiral dust filaments are externally heated and have a dust 
temperature of the order of 200~K (Cotera et al. 1999).
Only in the immediate vicinity of 
embedded stars (i.e. IRS~3, IRS~7 - at 0.5'' angular resolution these are the 
hottest dust sources known until now) 
the dust temperature is of the order of 200-300~K (see Cotera et al. 1999).
For the DSO we need 650~K (Gillessen et al. 2012a) to explain the K-band emission 
only by dust. So it would have to be exceptionally hot and/or have an exceptional 
overabundance of small grains.
At a temperature of only 450~K this scenario can be excluded (Fig.~\ref{fig1} right).
Applying the same surface brightness of more than 14$^m$/arcsec$^2$ for the DSO 
to the mini-spiral, it would be much brighter and clearly stand out in its dust emission at 2$\mu$m,
which is in contradiction to the observations.
The alternative is internal heating - however, that would require an internal heating source, i.e. most likely a star.
In Fig.~\ref{fig1}(right)
we show a decomposition in which we assume that 50\% of the current K$_s$-band flux
is due to a late dwarf (blue line), an A/F giant, or AGB star (magenta line).
We added to it dust at a temperature of 450~K (plotted in red)
The sum of the spectra is shown by the thick black line (dashed at short 
wavelengths for the AF giant/AGB case).
The points correspond 
to the L- and  K$_S$-band magnitudes, and H-band upper limit from Eckart et al. (2013a)
and the M-band measurement and K-band upper limit of Gillessen et al. (2012ab). 
Red and blue dashed curves also show their 550\,K and 650\,K warm dust fits. 
In solid blue, red, and magenta lines the emission from three different possible stellar 
types of the DSO core are plotted. Any of these stars embedded in 450\,K dust (solid brown line) 
can produce the black line that fits all the NIR DSO photometric measurements.
Black body luminosities and the detection of photospheric 
emission imply possible stellar luminosities of up to 30 \solar;
i.e. masses of 10-20 \solm ~are possible.
Details are given by Eckart et al. (2013a).

The simulations by Jalali et al. (2013, submitted) show that dusty sources like the DSO or the
IRS13N cluster can actually be formed from small molecular cloud complexes.
If they are on elliptical orbits (e.g. originating from the CND). The gravitational focusing
during the periapse passage close to the SgrA* black hole is capable to trigger star formation on time scales
required for young massive stars. This process actually helps forming stars in the vicinity of 
super massive black holes.

\section{\bf The G2 tail}
In Fig.2-3 by Eckart et al. (2013b) we compare the G2 tail emission to
8.6$\mu$m observation taken in 2004 with VISIR at the ESO VLT
 and in 1994 with the Palomar telescope (Stolovy et al. 1996).
These MIR results, the comparably low proper motion,  as well as the possible
linkage to the two mini-spiral disk systems, suggest that the 'tail' component
(red in Fig~\ref{fig2}c) might not be associated with the G2 source but rather is a back/fore-ground dust source 
associated to the mini-spiral within the central stellar cluster.
The long dust lane feature that crosses the mini-spiral to the south-east of the 
DSO (Fig.7 in Gillessen et al. 2013) may be a consequence of the 
interaction between the two rotation gas disk 
components that are associated with the northern and eastern arms.
Details are shown in our Fig~\ref{fig2}abc
(see also Fig.21 in Zhao et al. 2009, and Fig.10 in Vollmer \& Duschl 2000).
If upcoming observations can confirm that the 
'tail' component is not associated with the DSO, 
it does not need to be taken into account in future simulations.

While the DSO has a marginal extent in Br$\gamma$ line emission of about 30mas 
(Gillessen 2013b), it may be compared to other dusty sources in the field
(see Eckart et al. 2013a).
In fact, the DSO may be a
compact source comparable to the cometary shaped sources X3 and X7
(Muzic et al. 2010, see also Sabha et al. 2014 in prep.). 
Its smaller size compared to X3 and X7 can be explained
by the higher particle density within the accretion stream close to 
SgrA* (e.g. Shcherbakov \&  Baganoff 2010; see our Fig~\ref{fig2}de).
Its size also depends on how earlier passages close to SgrA* possibly
influence the distribution of gas and dust close to a possible 
star at the center of the DSO.

\section{Conclusions}
The observations planned for 2014 will be essential to investigate 
how the close flyby of the DSO/G2 object will alter the accretion characteristics of SgrA*.
The structural evolution of the DSO will also show if the DSO harbors a star or 
is a pure gas and dust cloud.
Further NIR imaging and spectroscopy data will also show if the extended 'tail' component is
associated with the head of the DSO/G2 source.
Theoretical investigations are required to study the formation process for DSO like sources
and how they are linked to the conditions of star formation in the central stellar cluster.
These studies will also have to address the question of whether the DSO is comparable to 
other dusty sources in the cluster - like the infrared 
excess IRS13N sources (Muzic  et al. 2008, Eckart et al. 2012b) or the 
bow-shock sources X3 and X7 (Muzic  et al. 2007, 2010).
Independent of the answers to these problems, the DSO flyby in 2014 will
undoubtedly be a spectacular event. It will reveal valuable information on the
physics in the immediate vicinity of super-massive black holes.


\end{document}